\documentclass[amsmath,superscriptaddress,showpacs,prb,twocolumn]{revtex4-1}
\usepackage{bbm}
\usepackage{bm}
\usepackage{enumerate}
\usepackage{graphicx}
\usepackage[dvips]{epsfig}
\usepackage{epsf}
\usepackage[latin1]{inputenc}
\usepackage{amsmath}
\usepackage{amsfonts}
\usepackage{amssymb}
\bibpunct{[}{]}{,}{n}{}{}

\begin{document}
	\title{Chiral topological insulator of magnons}
	\author{Bo Li}
	\affiliation{Department of Physics and Astronomy and Nebraska Center for Materials and Nanoscience, University of Nebraska, Lincoln, Nebraska 68588, USA}
	\author{Alexey A. Kovalev}
	\affiliation{Department of Physics and Astronomy and Nebraska Center for Materials and Nanoscience, University of Nebraska, Lincoln, Nebraska 68588, USA}
	\begin{abstract}
We propose a magnon realization of 3D topological insulator in the AIII (chiral symmetry) topological class. The topological magnon gap opens due to the presence of Dzyaloshinskii-Moriya interactions. The existence of the topological invariant is established by calculating the bulk winding number of the system.  Within our model, the  surface magnon Dirac cone is protected by the sublattice chiral symmetry.  By analyzing the magnon surface modes, we confirm that the backscattering is prohibited. By weakly breaking the chiral symmetry, we observe the magnon Hall response on the surface due to opening of the gap. Finally, we show that by changing certain parameters the system can be tuned between the chiral topological insulator, three dimensional magnon anomalous Hall, and Weyl magnon phases.
	\end{abstract}
	\maketitle

	\section{Introduction}
	
The discovery of topological insulators (TIs) \cite{RevModPhys.82.3045,RevModPhys.83.1057} is a remarkable achievement in condensed matter physics as it reveals fundamental connection to topology and is promising for applications in electronics and quantum computing. At the same time, the concept of topology arises in a variety of other fields under the encouragement of the success of topological insulators \cite{Khanikaev.HosseinMousavi.ea:NM2013,Suesstrunk.Huber:P2016}. Recently, there has been considerable interest in the topological physics of magnon systems \cite{Katsura.Nagaosa.ea:PRL2010,Mook.Henk.ea:PRB2014,PhysRevB.89.134409,Mook.Henk.ea:PRB2015,Lee.Han.ea:PRB2015,Kim.Ochoa.ea:PRL2016,Owerre:JAP2016,2017arXiv171208170H,2018arXiv180204283M}. Realizations of systems with a Weyl spectrum of magnons have been suggested \cite{Li.Li.ea:NC2016,Mook.Henk.ea:PRL2016,Su.Wang.ea:PRB2017,PhysRevB.96.104437,PhysRevB.97.094412, PhysRevB.97.115162}. Multiple theoretical works \cite{PhysRevLett.95.057205,PhysRevB.87.174427,PhysRevB.87.174402,PhysRevB.87.144101,Mook.Henk.ea:PRB2014,PhysRevLett.115.147201,PhysRevB.89.134409,0953-8984-28-38-386001,PhysRevB.93.214403,PhysRevLett.118.177201,PhysRevB.95.014435,2017arXiv170609548W,PhysRevLett.117.217203,Owerre:JAP2016,Nakata.Kim.ea:PRB2017, PhysRevB.97.174407,1367-2630-18-4-045015,PhysRevB.97.081106,2018arXiv180109945P,PhysRevB.97.180401} have discussed the edge or surface states of gapped magnon systems. Due to the absence of the Kramers degeneracy and the electronic orbital freedom for magnons, the investigation has been limited to the magnon analog of the Chern insulator. A magnon analog of the quantum spin Hall effect comprised of two copies of magnon Chern insulators has also been proposed \cite{PhysRevLett.117.217203,Nakata.Kim.ea:PRB2017}. Nevertheless, the topological protected helical surface states have not been discussed for magnon systems. According to the ten-fold way classification of TIs \cite{PhysRevB.78.195125,1367-2630-12-6-065010}, the AIII class only requires the sublattice chiral symmetry for realization of a topological insulator with $\mathbb{Z}$ invariant in one and three dimensions \cite{PhysRevB.81.045120,PhysRevLett.113.033002,HASEBE2014681,PhysRevB.95.121107}. Hosur \textit{et al.} \cite{PhysRevB.81.045120} discussed an electronic model of chiral topological insulator (cTI). Wang \textit{et al.} suggested a realization of cTI in cold-atom systems \cite{PhysRevLett.113.033002}. 
	
\begin{figure}\centerline{\includegraphics[width=1\columnwidth]{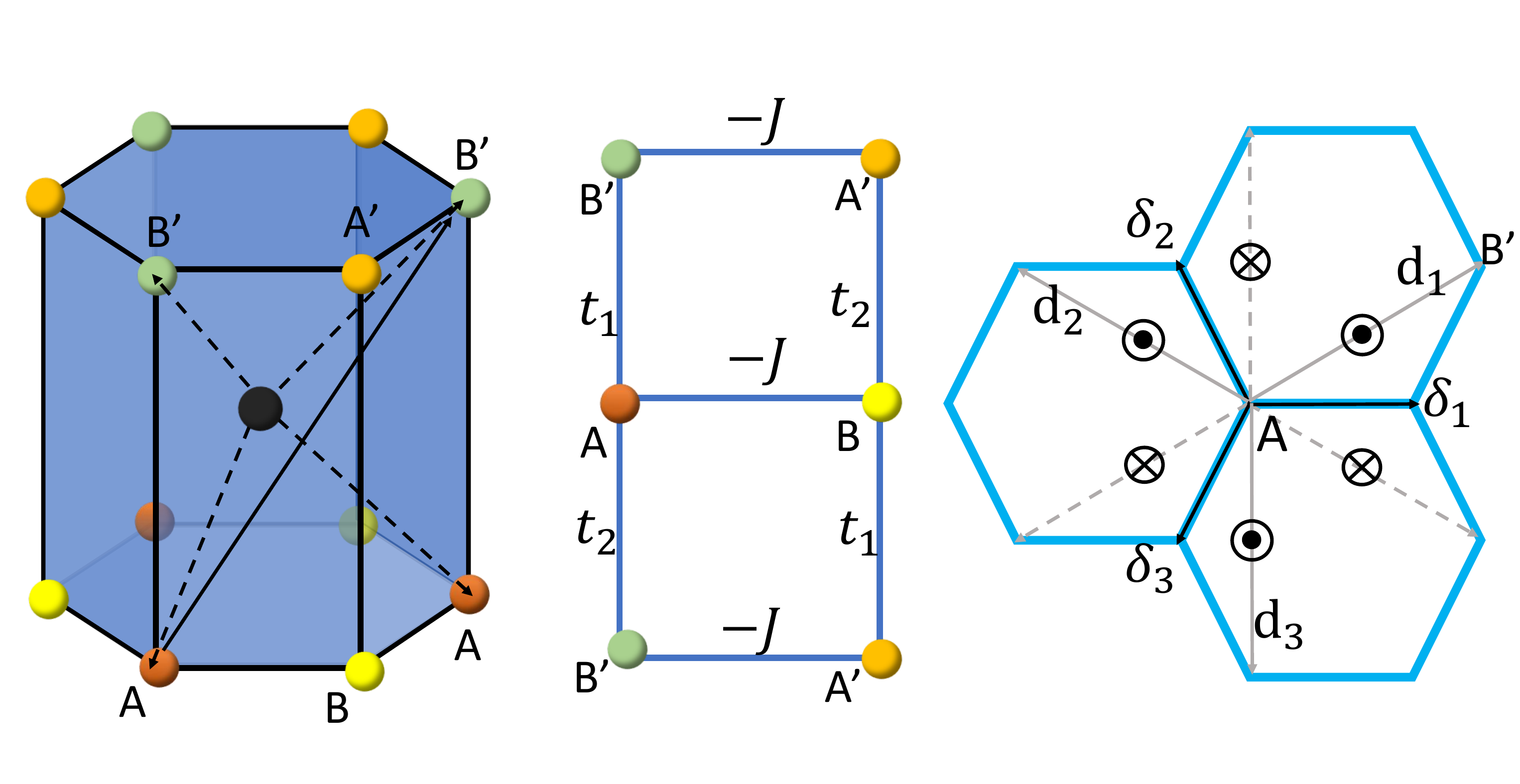}}
\protect\caption{(Color online) Left: The layered Honeycomb structure. The central non-magnetic atom generate DMI between interlayer third-nearest-neighbor atoms, e.g., $A$ and $B^\prime$. Middle: The in-plane and interlayer exchange energy. Right: The projection of interlayer DMI between $A$ and $B^\prime$ on $z$ direction. }
\label{figure1}
\end{figure}

In this paper, we show that magnon chiral topological insulator (mcTI) can be realized in a Heisenberg model endowed with the Dzyaloshinskii-Moriya interaction (DMI) \cite{Dzyaloshinsky:JPCS1958,Moriya:PR1960}. We consider a layered honeycomb lattice structure \cite{PhysRevB.94.075401,PhysRevX.8.011010} in which the interactions are chosen such that the system possesses the chiral symmetry (see Fig.~\ref{figure1}). The bulk is characterized by the $\mathbb{Z}$ topological invariant: winding number. In accordance with the bulk-boundary correspondence, our model supports a symmetry-protected magnon Dirac cone on its surface, provided the chiral symmetry is not broken on the surface.  The helical surface states lack backscattering in the presence of the chiral symmetry. By breaking the chiral symmetry, a small gap can be introduced in surface band, which leads to the magnon Hall response, e.g., under a temperature gradient. We observe that similar to electronic systems, the chiral symmetric perturbations can change the system to the nodal line and trivial phases. Furthermore, by adding terms breaking the chiral symmetry, we can bring our system into the three-dimensional magnon anomalous Hall (3D-mAH), and Weyl magnon phases.   	

The paper is organized as follows. In Sec. II (and in Appendix B), we construct models of mcTI and clarify the presence of the chiral symmetry and the mass term. In Sec. III, we calculate the topological invariant associated with the spectrum of magnons in mcTI. In Sec. IV, we study the surface states by constructing the effective Hamiltonian and calculating the Hall-like response to the temperature gradient. In Sec. V, we vary various parameters of the model and construct a phase diagram with the nodal line and mcTI phases. Several appendices give more details about our calculations. 
	
\section{Model}
	
We consider a layered honeycomb magnetic structure with ferromagnetic ordering, as shown in Fig.~\ref{figure1}. To realize mcTI, we construct a model with the magnon Dirac spectrum in the bulk. We then open a gap by adding a mass term corresponding to additional DMI. In Appendix B, we show that there are various ways to introduce the mass term. 
The Heisenberg Hamiltonian is composed of the in-plane and interlayer exchange interactions, and the axial anisotropy terms,		
	\begin{equation}\label{eq1}
	H=H_{\text{in}}+H_{\text{inter}}+H_{\text{an}},
	\end{equation}
where
	\begin{eqnarray}\label{eq2}
	H_{\text{in}}&=&-J\sum_{z,i}\sum_{\mu=1}^{3} \mathbf{S}_{A,i}\cdot\mathbf{S}_{B,i+\bm{\delta}_\mu}+\mathbf{S}_{A^\prime,i}\cdot\mathbf{S}_{B^\prime,i-\bm{\delta}_\mu},\nonumber\\
	H_{\text{inter}}&=-&\sum_{i,z}(t_1\mathbf{S}_{A,z}\cdot\mathbf{S}_{B^\prime,z+1}+t_2\mathbf{S}_{A,z}\cdot\mathbf{S}_{B^\prime,z-1})\nonumber\\
	&&+(t_1\leftrightarrow t_2, A\rightarrow B,B^\prime\rightarrow A^\prime),\nonumber\\
	H_{\text{an}}&=&\sum_{i,z}\sum_{Q}K(S_{Q,(i,z)}^z)^2.
	\end{eqnarray}
Here $i$ corresponds to the in-plane index and $z$ corresponds to the layer index; $\bm{\delta}_1=(1,0,0)$,  $\bm{\delta}_2=(-\frac{1}{2},\frac{\sqrt{3}}{2},0)$, $\bm{\delta}_3=(-\frac{1}{2},-\frac{\sqrt{3}}{2},0)$; $J$ and $K$ are nearest exchange and axial anisotropy energy with $K<0$. $Q$ stands for different spin modes, i.e., $Q=A,B,A^\prime,B^\prime$. In the Hamiltonian, we suppress unrelated coordinates for clearness. For in-plane interaction, we only consider nearest-neighbor exchange. For the interlayer interaction, we use a staggered pattern as shown in Fig.~\ref{figure1} (this limitation simplifies analysis but it is not necessary, as we show in Sec.~V). We perform Holstein-Primakoff transformation in the large $S$ limit, $S_{Q,i}^z=(S-Q_i^\dagger Q_i)$ and $S_{Q,i}^+=\sqrt{2S}Q_i$, with $Q_i^\dagger$, $Q_i$ being the magnon creation and annihilation operators for spin mode $S_{Q}$. The Hamiltonian in momentum  space is written in the basis $\Psi_\mathbf{k}=(A_\mathbf{k},B_\mathbf{k},A^\prime_\mathbf{k},B^\prime_\mathbf{k})$, where we label the layer and sublattice degrees of freedom by $\mu$ and $\tau$ Pauli matrices, 
	\begin{equation}\label{eq3}
	H=JS\sum_{\mathbf{k}}\Psi_\mathbf{k}^\dagger\mathcal{H}_\mathbf{k}\Psi_\mathbf{k},
	\end{equation}
with
	\begin{eqnarray}\label{eq4}
	\mathcal{H}_\mathbf{k}&=&\varepsilon_0-\gamma_{1\mathbf{k}}\tau_x+\gamma_{2\mathbf{k}}\mu_z\tau_y+ 2\lambda \cos(k_z)\mu_x\tau_x\nonumber\\
	&&-2\delta \sin(k_z)\mu_x\tau_y.
	\end{eqnarray}
Here $\varepsilon_0=3-2\lambda-2\kappa$, $\gamma_\mathbf{k}=\sum_\mu e^{i\mathbf{k}\cdot\bm{\delta}_\mu}=\gamma_{1\mathbf{k}}+i\gamma_{2\mathbf{k}}$, with $\gamma_{1\mathbf{k}}=\cos(k_x)+2\cos(\frac{k_x}{2})\cos(\frac{\sqrt{3}k_y}{2})$ and $\gamma_{2\mathbf{k}}=2[\cos(\frac{k_x}{2})-\cos(\frac{\sqrt{3}k_y}{2})]\sin(\frac{k_x}{2})$, $\lambda=-\frac{1}{2}(t_1+t_2)/J$,  $\delta=\frac{1}{2}(t_2-t_1)/J$, and $\kappa=K/J$. Note that the Hamiltonian above has the chiral symmetry $\tau_z$ up to a constant term  (below, we disregard this constant energy shift), i.e.,
	\begin{equation}\label{eq5}
	\tau_z\mathcal{H}_\mathbf{k}\tau_z=-\mathcal{H}_\mathbf{k}.
	\end{equation}
First, we consider the case $\lambda=0$, corresponding to the staggered interlayer exchange. In this pattern, the exchange term realizes the so-called $\pi$ flux \cite{PhysRevB.81.045120} for vertical plaquettes $\Pi_\square sign(t_{ij})=-1 $, e.g.,  $AB^\prime A^\prime BA$, where $t_{ij}$ stands for the exchange strength between two spins. The eigenenergy,
	\begin{equation}\label{eq6}
	E_\pm/JS=\pm\sqrt{|\gamma_{\mathbf{k}}|^2+4\delta^2 \sin^2(k_z)},
	\end{equation}
reveals two Dirac cones at $\mathbf{Q}_{R(L)}=(0,\pm \frac{4\pi}{3\sqrt{3}},0)$. Around the Dirac point $\mathbf{Q}_R$, the Hamiltonian reads 
	\begin{equation}\label{eq7}
	\mathcal{H}_{0,\mathbf{k}}=q_i\alpha_i,
	\end{equation}
where $q_x=\frac{3}{2}k_x$, $q_y=\frac{3}{2}k_y$, and $q_z=-2\delta k_z$; $\{\alpha_i\}=\{\mu_z\tau_y,\tau_x, \mu_x\tau_y\}$ satisfy the relation $\{\alpha_i,\alpha_j\}=2\delta_{ij}$. For the other Dirac point, the Hamiltonian is easily obtained after the transformation $q_y\rightarrow -q_y$ in Eq.~\eqref{eq7}. Since the two Dirac cones give us equivalent physics, we use the form in Eq.~\eqref{eq7} in the following discussion.
	
To realize mcTI, the Hamiltonian should have a chiral symmetric mass term to open the gap in the bulk Dirac cone while preserving the surface Dirac cone. In a massive Dirac equation for the bulk, the mass term is described by the matrix $\beta$ satisfying the anti-communication relation $\{\beta, \mathcal{H}_{0,\mathbf{k}}\}=0$. The only possible term preserving the chiral symmetry is  $\beta=\mu_y\tau_y$.  To this end, we include the third-nearest-neighbor interlayer DMI in our model. The correct form of DMI can be produced by the central non-magnetic atom as it is shown in Fig.~\ref{figure1}, where we assume an overlap of relevant orbitals and a sufficiently strong spin-orbit interaction. The DMI term becomes  	
	\begin{eqnarray}\label{eq8}
	H_{\text{th}}&=& \sum_{z,n=\pm 1}\sum_{i,\mathbf{d}_\lambda}\mathbf{D}_{AB^\prime}(\mathbf{d}_\lambda)\cdot[\mathbf{S}_{A,(i,z)}\times\mathbf{S}_{{B^\prime},(i+\mathbf{d}_\lambda,z+n)}]\nonumber\\
	&&+\{A\rightarrow B, B^\prime\rightarrow A^\prime\},
	\end{eqnarray}
where $i$, $z$ are the in-plane and layer coordinates with assumption of unit interlayer distance in $z$ direction, $\mathbf{d}_\lambda$ represents the in-plane second-nearest-neighbor between atoms with $\mathbf{d}_1=(\frac{3}{2},\frac{\sqrt{3}}{2},0)$, $\mathbf{d}_2=(-\frac{3}{2},\frac{\sqrt{3}}{2},0)$, and $\mathbf{d}_3=(0,-\sqrt{3},0)$ (the other three are $-\mathbf{d}_1$,$-\mathbf{d}_2$,$-\mathbf{d}_3$). At the same time, we assume that the in-plane DMI between the second-nearest-neighbors is absent, as such a term would break the chiral symmetry. For the magnetization along the $z$ axis, only the $z$ component of DMI vectors is relevant, which is shown in Fig.~\ref{figure1}. The $z$ projections of DMI vectors have the same magnitude $D^z$ and follow the staggered pattern shown in Fig.~\ref{figure1}.    In momentum space, the DMI term reads
\begin{equation}\label{eq9}
\mathcal{H}_{\text{th}}=4\delta_D \xi_\mathbf{k}\cos(k_z)\mu_y\tau_y,
\end{equation}
where $\delta_D=D^z/J$ and $\xi_\mathbf{k}=\sum_{i=1}^3\sin(\mathbf{k}\cdot \mathbf{d}_i)$. Now, we have the full model given by Eqs.~\eqref{eq4} and \eqref{eq9}.

\begin{figure}[h]
			\centering
			\includegraphics[width=1\columnwidth]{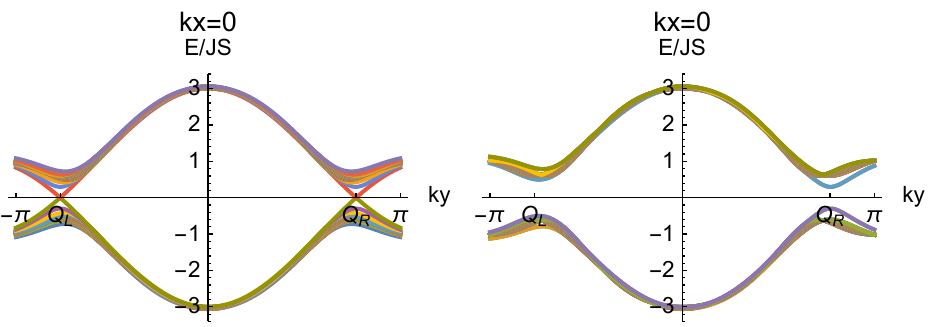}
			\protect\caption{(Color online) A plot corresponding to a slab geometry with the parameters, $\delta=0.3$, $\delta_D=0.15$. Left: The surface state with the Dirac cone at $\mathbf{Q}_L$ and $\mathbf{Q}_R$ where the surface states appear at the interface between the two bulk regions with the opposite sign of DMI $\delta_D$. Right: The surface state cone splits when the bulk is interrupted at a honeycomb plane in contact with vacuum due to uncompensated exchange interactions leading to breaking of the chiral symmetry.}\label{figure2}
\end{figure}

To confirm the existence of surface states, we diagonalize the Hamiltonian given by Eqs.~(\ref{eq1}) and (\ref{eq8}) in a slab geometry. In our calculation, we consider two bulk regions with the opposite sign of DMI $\delta_D$, which guarantees the sign change of the mass term across the interface.  As expected, the model has Dirac states confined to the $x-y$ plane separating the two bulk regions as shown in Fig.~\ref{figure2}, left. The model hosts two surface Dirac cones at the two-dimensional projection of $\mathbf{Q}_R$ and $\mathbf{Q}_L$ as long as all parameters are nonzero. We also considered a bulk terminated at a honeycomb plane with vacuum,  which results in a single Dirac cone with a gap opening due to breaking of the chiral symmetry at the interface (see Fig.~\ref{figure2}, right).	The chiral symmetry breaking appears due to the exchange energy terms at the interface after application of the Holstein-Primakoff transformation.	

	\section{topological invariant}

The presence of chiral symmetry ensures that the Hamiltonian could be brought to an off-diagonal form by a unitary transformation. For our case, we need a transformation satisfying $U\tau_zU^\dagger=\mu_z$, under which,
	\begin{eqnarray}\label{eq10}
	\tilde{\mathcal{H}}_{\mathbf{k}}=U\mathcal{H}_{\mathbf{k}} U^\dagger= \left[
	\begin{array}{cc}
	0&D_\mathbf{k}\\
	D_\mathbf{k}^\dagger&0
	\end{array}\right],
	\end{eqnarray}
with 
	\begin{eqnarray}\label{eq11}
	D_{\mathbf{k}}=\left[
	\begin{array}{cc}
	-\gamma_{\mathbf{k}}&\Delta_\mathbf{k}\\
	-\Delta_\mathbf{k}^\ast&-\gamma_{-\mathbf{k}}\\
	\end{array}\right],
	\end{eqnarray}
where $\Delta_{\mathbf{k}}=-4\delta_D\xi_\mathbf{k}\cos(k_z)+i 2\delta \sin(k_z)$. We can adiabatically deform $\tilde{\mathcal{H}}_{\mathbf{k}}$ into a flat-band Hamiltonian $Q_\mathbf{k}=1-2\sum_{a\in \text{B.G.}}|\psi_a\rangle\langle\psi_a|$ \cite{PhysRevB.78.195125,1367-2630-12-6-065010} where $\psi_a$ is the eigenstate of $\tilde{\mathcal{H}}_{\mathbf{k}}$ and $\text{B.G.}$ stands for the states below the gap. The matrix form reads
	\begin{eqnarray}\label{eq12}
	Q_\mathbf{k}=\left[\begin{array}{cc}
	0&q_\mathbf{k}\\
	q^\dagger_\mathbf{k}&0
	\end{array}\right],
	\end{eqnarray}
where the off-diagonal term is $q_\mathbf{k}=\frac{1}{\lambda}D_\mathbf{k}$ with $\lambda=\sqrt{|\gamma_\mathbf{k}|^2+|\Delta_\mathbf{k}|^2}$.
	The chiral topological state can be characterized by the three-dimensional winding number \cite{PhysRevB.78.195125,1367-2630-12-6-065010}
	\begin{equation}\label{eq13}
	\nu[q]=\int\frac{d^3k}{24\pi^2}\epsilon^{\mu\nu\rho}\text{tr}[(q^\dagger\partial_\mu q)(q^\dagger\partial_\nu q)(q^\dagger\partial_\rho q)],
	\end{equation}
where $\mu,\nu=k_x,k_y,k_z$ and the integration goes over the whole Brillouin zone. Numerical results show that the winding number is quantized for nonzero $\delta_D$ and $\delta$. When $\delta_D=0$ or $\delta=0$, the model falls into the Dirac phase with vanishing winding number.  
This result can be understood (details in Appendix~C) by considering the topologically equivalent Hamiltonian around $\mathbf{Q}_{R}$: $\mathcal{H}_{\mathbf{Q}_R+\mathbf{k}}=q_i\alpha_i+m\mu_y\tau_y$ with $m=2\sqrt{3}\delta_D$ (here we drop the momentum dependence of mass term in topological sense). The topological invariant is calculated as $\nu_R[q]= \text{sgn}(\delta_D\delta)/2$. For $\mathbf{Q}_L$ point, we replace $q_y\rightarrow-q_y$ and $m\rightarrow-m$ to get $\nu_L[q]=\text{sgn}(\delta_D\delta)/2$. The total winding number is the sum,
	\begin{equation}\label{eq14}
	\nu[q]=\text{sgn}(\delta_D\delta),
	\end{equation} 
which is a quantized number for the nontrivial mcTI phase and zero for the trivial phase. In our model, there is only one  Dirac cone on the surface projection point of $\mathbf{Q}_R$ or $\mathbf{Q}_L$. Specifically, when $\nu[q]=1 (-1)$, the Dirac cone appears on the projection of $\mathbf{Q}_R$ ($\mathbf{Q}_L$) point. In general, mcTI can have more than one Dirac cone at the boundary.

\section{Surface state}

\begin{figure}[h]
		\centering
		\includegraphics[width=0.8\columnwidth]{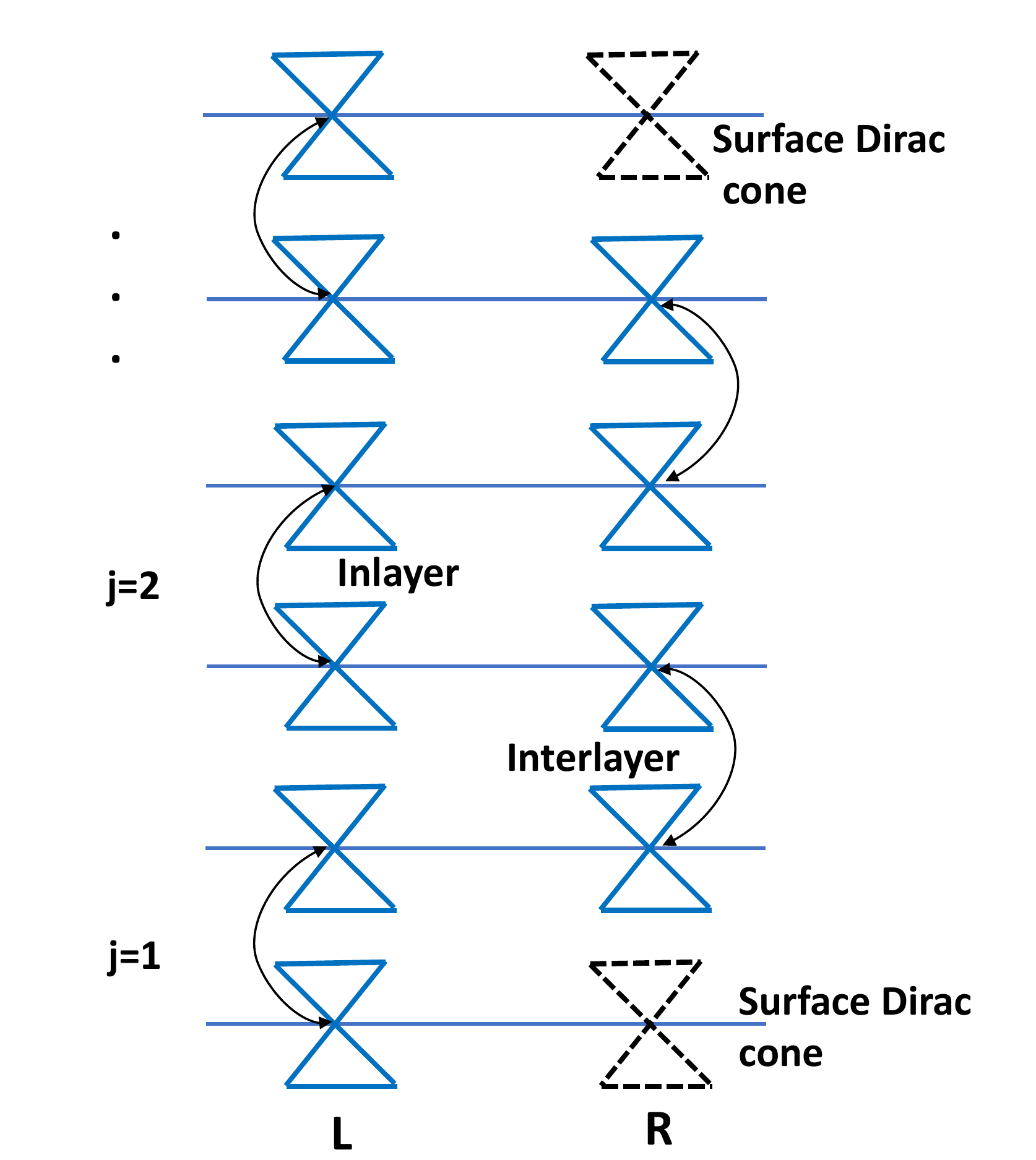}
		\protect\caption{(Color online) Pairing pattern for $\delta=\frac{\sqrt{3}}{2}\delta_D$. The $R$ Dirac cone resides on the surface.}\label{figure4}
	\end{figure}

We can get a physical insight into the formation of the surface Dirac cone by considering the interlayer Dirac cone pairing pattern \cite{PhysRevB.81.045120}. 
For simplicity, we ignore the chiral symmetry-breaking terms appearing when we terminate a sample at one of honeycomb planes in contact with vacuum.
Such symmetry-breaking terms do not appear if the interface is formed between the two bulk regions with the opposite sign of DMI $\delta_D$ or if the interface is terminated in such a way that the chiral symmetry-breaking terms due to exchange energy do not appear.
We consider the Hamiltonian that is Fourier transformed with respect to the in-plane momentum,
\begin{eqnarray}\label{eq15new}
	&H_{j,j}=-\gamma_{1\mathbf{k}_\parallel}\tau_x+\gamma_{2\mathbf{k}_\parallel}\mu_z\tau_y+(2\xi_{\mathbf{k}_\parallel}\delta_D-\delta)\mu_y\tau_y,\nonumber\\
	&H_{j,j\pm1}=\pm i(\delta+2\delta_D\xi_{\mathbf{k}_\parallel})\mu_\mp\tau_y,
	\end{eqnarray}
where the index $j$ labels the bilayer, $H_{j,j}$ describes intralayer terms, and $H_{j,j\pm1}$ describes the interlayer terms in the Hamiltonian written in the basis $(A_{\mathbf{k}_{\parallel},j},B_{\mathbf{k}_{\parallel},j},A^\prime_{\mathbf{k}_{\parallel},j},B^\prime_{\mathbf{k}_{\parallel},j})$, with $\mathbf{k}_{\parallel}$ representing the in-plane momentum (see Fig.~\ref{figure4}).
The intralayer Hamiltonians describe two-dimensional Dirac cones (different from the bulk Dirac cones discussed before), which hybridize due to interlayer coupling. It is convenient to consider the Hamiltonian written in the subspace $(A_{R,j},B_{R,j},A_{R,j}^\prime,B_{R,j}^\prime,A_{L,j},B_{L,j},A_{L,j}^\prime,B_{L,j}^\prime)$ where index $R(L)$ stands for the in-plane momentum $(0,\pm\frac{4\pi}{3\sqrt{3}})$, and Pauli matrix $\nu_z$ acts on $R$ and $L$ Dirac cones,
	\begin{eqnarray}\label{eq15}
	&H_{j,j}=-(\delta-\sqrt{3}\delta_D\nu_z)\mu_y\tau_y,\nonumber\\
	&H_{j,j\pm1}= \pm i(\delta+\sqrt{3}\delta_D\nu_z)\mu_\mp\tau_y.
	\end{eqnarray}
Here $\mu_\pm=\frac{1}{2}(\mu_x\pm i\mu_y)$. For $\delta=\sqrt{3}\delta_D$, we obtain that $H_{j,j}\propto \frac{1-\nu_z}{2}$ and $H_{j,j\pm1}\propto \frac{1+\nu_z}{2}$, which shows that $R$ and $L$ Dirac cones hybridize in a pattern shown in Fig.~\ref{figure4}. In this special case, the surface states live on top and bottom surfaces without any penetration into the bulk.  If $\delta=-\sqrt{3}\delta_D$, the $R$ and $L$ cones interchange in the hybridization pattern.
We can investigate the surface states further in the vicinity of $(0,\pm\frac{4\pi}{3\sqrt{3}})$ point using the $\mathbf{k}\cdot\mathbf{p}$ theory. After replacing $k_z$ to its second order by $-i\partial_z$  in the Hamiltonian, the effective Hamiltonian becomes 
\begin{equation}\label{eq16}
H(z)=-2\delta(-i\partial_z)\mu_x\tau_y+M[1-\frac{1}{2}(-i\partial_z)^2]\mu_y\tau_y,
\end{equation}
with $M=4\delta_D\xi_{\mathbf{Q}_{R(L)}} (=\pm 2\sqrt{3}\delta_D)$.  Under the boundary condition that the wave function vanishes at $z=0$ and $z=\infty$, and taking the same termination as above, we obtain the eigenstates for the Hamiltonian (see Appendix~D) as below,
\begin{eqnarray}\label{eq17}
\psi_{1}(z)&=&f(z)(0,1,0,0)^T,\nonumber\\
\psi_{2}(z)&=& f(z)(1,0,0,0)^T.
\end{eqnarray}
Here $f(z)=\sqrt{2(1-\frac{2}{\beta^2})}e^{-\beta z}\sinh(\sqrt{\beta^2-2}z)$, $\beta=2\delta/M$, and $\beta>0$ has to be satisfied to ensure the existence of the surface state. For a given $\delta$, $\beta_R=-\beta_L$ with $\beta_{R(L)}$ being the value of $\beta$ at $\mathbf{Q}_{R(L)}$ point. It is clear that $\beta_R>0$ when $\text{sgn}(\delta_D\delta)>0$, and the surface Dirac cone exists at the projection of $\mathbf{Q}_R$ point; when $\text{sgn}(\delta_D\delta)<0$,  $\beta_L>0$, and the surface Dirac cone exists at the projection of  $\mathbf{Q}_L$ point. This result is consistent with the earlier discussion.

Without loss of generality, we consider the surface state existing at the projection of $\mathbf{Q}_R$ point. The effective Hamiltonian is
\begin{equation}\label{eq19}
H_{\text{sur}}=v_F(\mathbf{k}\times \mathbf{e}_z)\cdot\bm{\tau},
\end{equation}
where $v_F=\frac{3}{2}$. This Hamiltonian exhibits magnon spin-momentum locking \cite{Okuma:PRL2017} in the spin space defined by sublattices $A$ and $B$. The Rashba-like surface states in Eq.~(\ref{eq19}) are described by helical eigenvectors, i.e., the eigenstate of $\mathbf{k}$ and $-\mathbf{k}$ are orthogonal to each other, which prohibits backscattering between states with opposite momentum. The chiral symmetric perturbation can only shift the position of the Dirac cone as it adds additional terms of the form $M_1\tau_x+M_2\tau_y$ to Eq.~({\ref{eq19}}). This is a manifestation of the fact that the surface modes are protected by chiral symmetry. \\

Interesting physics can also arise when the chiral symmetry is weakly broken at the interface.
We can break the surface Dirac cone by considering an interface with vacuum (see Fig.~\ref{figure2}) or by contacting mcTI with another material that has a broken chiral symmetry. The gapped effective surface Hamiltonian reads,  $\mathcal{H}_{\text{sur}}=v_F(\mathbf{k}\times \mathbf{e}_z)\cdot\bm{\tau}+m_s\tau_z$. The gap in the surface Dirac cone will result in a Hall response to a longitudinal driving force on the surface, similar to the surface Hall effect in 3D topological insulators with broken time-reversal symmetry  \cite{PhysRevB.84.085312}, which can be detected by the spin Nernst response \cite{PhysRevB.93.161106}, 
\begin{eqnarray}\label{eq20}
j_y^s=\alpha_{yx}\nabla_x T,
\end{eqnarray}
with response parameter $\alpha_{yx}=-\frac{k_B}{V}\sum_{\mathbf{k},n}\Omega_{yx}^n(\mathbf{k})c_1(g(\varepsilon_n))$, where $V$ is the surface area of the system, $\Omega_{yx}^n(\mathbf{k})$ is the momentum space Berry curvature, $c_1(x)=(1+x)\ln(1+x)-x\ln x$, $g(\varepsilon)=1/(e^{\beta \varepsilon}-1)$ is the Boson-Einstein distribution function (see Appendix~D). To identify the contribution from the Dirac cone, we introduce a cutoff $\Lambda$ such that $\Lambda < \varepsilon_0$. The response parameter is calculated as
\begin{eqnarray}\label{eq21}
\alpha_{yx}&\approx&\frac{\pi k_Bm_s\varepsilon_0\beta^2}{\cosh(\beta\varepsilon_0)-1} \ln(\Lambda/|m_s|),
\end{eqnarray} 
where $\beta=JS/(k_BT)$. Unlike electronic system, the response parameter is not quantized due to the Bose-Einstein statistics. In Eq.~(\ref{eq21}), only the contribution from the Dirac cone has been considered. We note that the Berry curvature from other parts of the Brillouin zone can also contribute to the spin Nernst response due to the Bose-Einstein statistics.

\begin{figure}
	\begin{tabular}{cc}
		{\includegraphics[width=0.5\columnwidth]{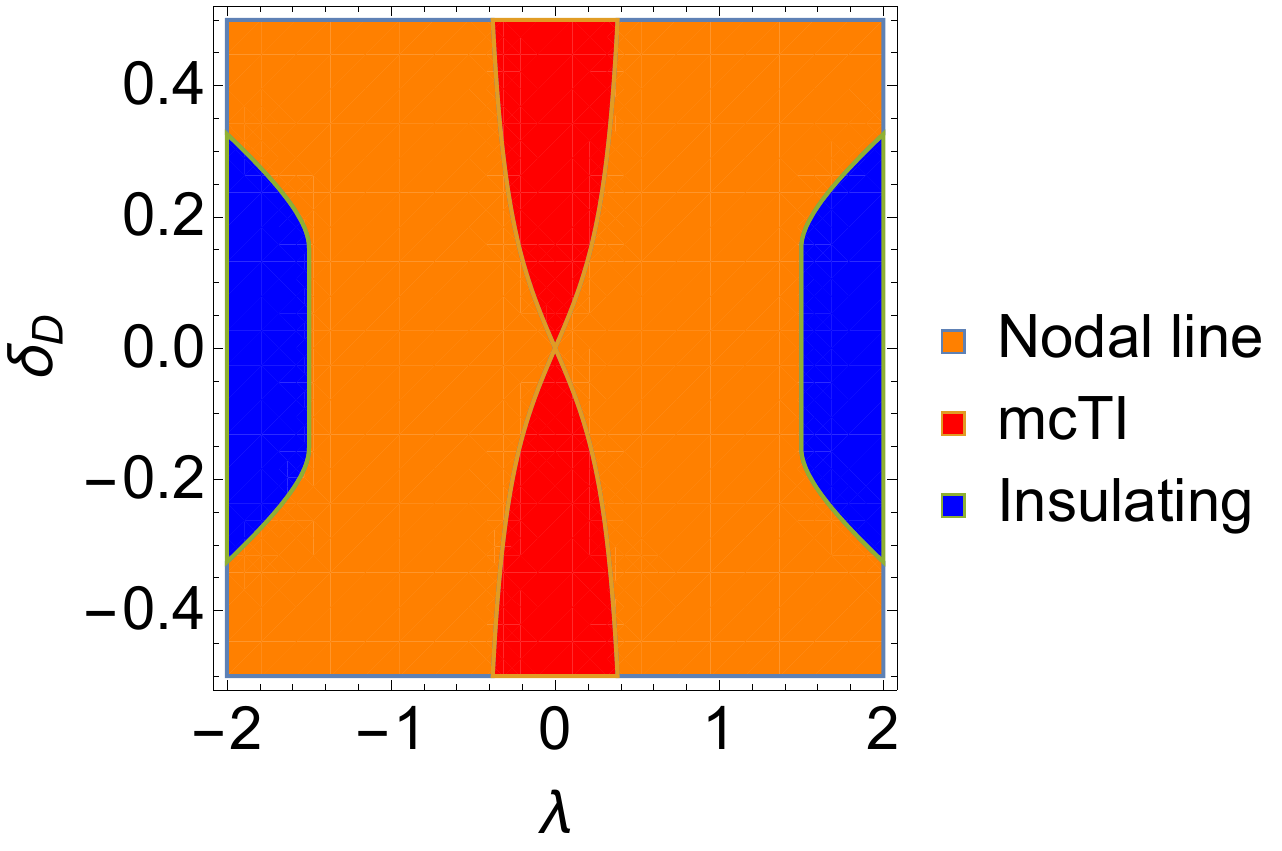}}&
		{\includegraphics[width=0.5\columnwidth]{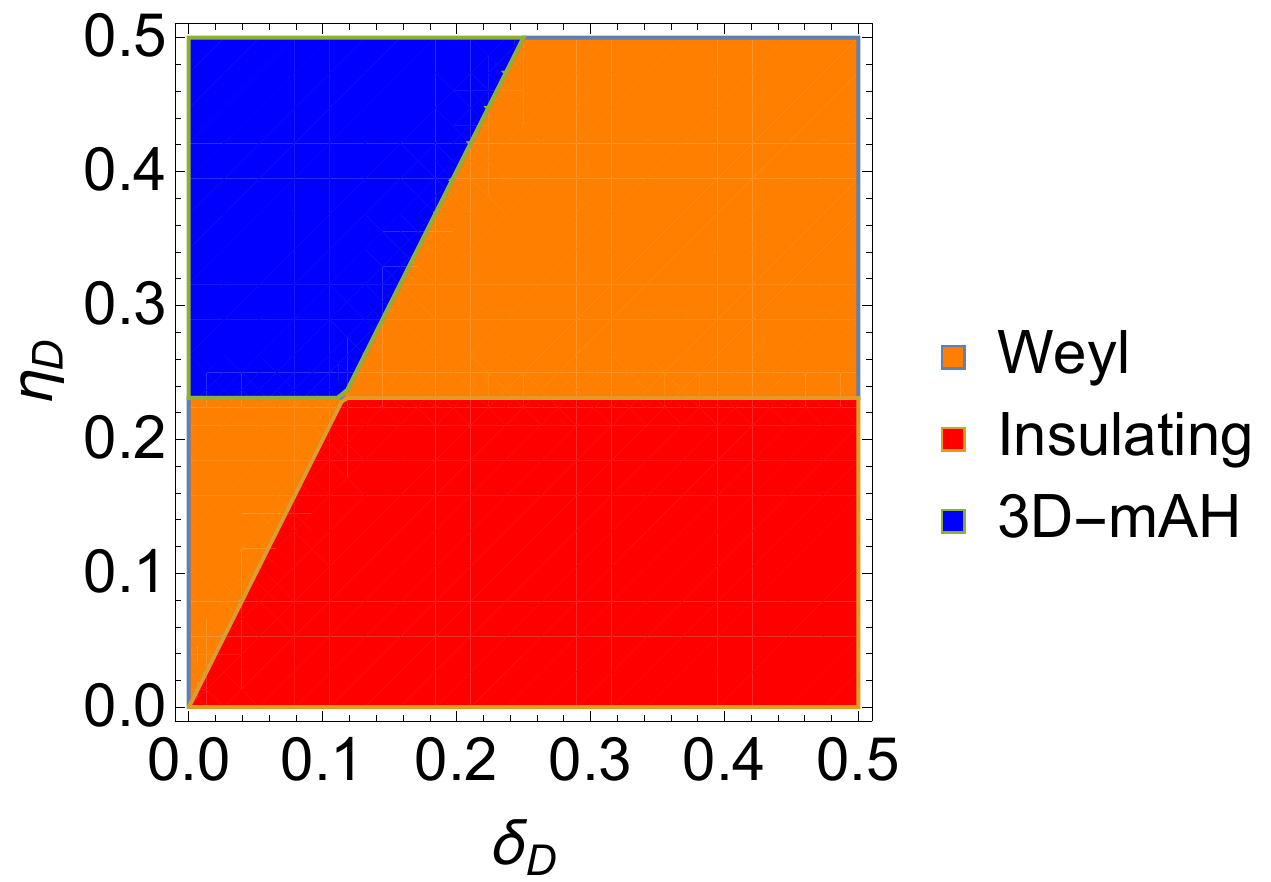}}
	\end{tabular}
	\protect\caption{(Color online) Left: Phase diagram in $ \delta_D-\lambda$ parameter space with $\delta\neq 0$. The mcTI phase is continuously connected to the $\lambda=0$ case considered in the previous sections. Right: Phase diagram in $\eta_D-\delta_D$ parameter space ($\lambda=0$)  with $\delta=0.2$; The boundary lines between different phases are $\eta_D=2\delta/\sqrt{3}$ and $\delta_D=\eta_D/2$.}
	\label{figure5}
\end{figure}

 \section{topological phase transition}
We now consider a more general model with a non-staggered pattern, i.e., $\lambda\neq 0$.  
We find that even for $\lambda\neq 0$ there is still some region in parameter space with mcTI phase. As we increase $\lambda$, we encounter a phase transition into a nodal line phase before we reach the trivial insulating phase (see Fig.~\ref{figure5}). For the full Hamiltonian composed of Eqs.~\eqref{eq4} and \eqref{eq9}, the energy is
 \begin{eqnarray}\label{eq22}
 E_\mathbf{k}^2/(JS)^2&=&[|2\lambda\cos k_z|\pm\sqrt{|\gamma_\mathbf{k}|^2+(4\delta_D)^2\xi_\mathbf{k}^2\cos^2k_z}]^2\nonumber\\
 &&+(2\delta)^2\sin^2k_z.
 \end{eqnarray}
To get nodal line phase, it's required that 
 $k_z=0$ and
 $(2\lambda)^2=|\gamma_\mathbf{k}|^2+(4\delta_D)^2\xi_\mathbf{k}^2$. When $\min\{|\gamma_\mathbf{k}|^2+(4\delta_D)^2\xi_\mathbf{k}^2\}\leqslant (2\lambda)^2\leqslant \max\{|\gamma_\mathbf{k}|^2+(4\delta_D)^2\xi_\mathbf{k}^2\}$, the system falls into the nodal line phase with the nodal lines lying on $k_z=0$ plane. When $(2\lambda)^2< \min\{|\gamma_\mathbf{k}|^2+(4\delta_D)^2\xi_\mathbf{k}^2\}$, it's in mcTI phase that is continuously related to the $\lambda=0$ case considered in the previous sections. Note that if $\delta=0$, the gap is always closed at $(0,\pm \frac{4\pi}{3\sqrt{3}},\pm\frac{\pi}{2})$, so that $\delta\neq 0$ has to be satisfied. The phase diagram is shown  in Fig.~\ref{figure5}. We find that there is a substantial region in parameter space with mcTI phase.

  Besides the phase transition induced in the presence of the chiral symmetry, we find that the system can also be tuned to the Weyl and 3D-mAH phase by introducing the in-plane second-nearest-neighbor bulk DMI that breaks the chiral symmetry,
  \begin{eqnarray}\label{eq23}
  \delta H=\frac{1}{2}\sum_{Q}\sum_{z,i,\mathbf{d}_\lambda}\tilde{D}_Q^z(\mathbf{d}_\lambda)\mathbf{e}_z\cdot[\mathbf{S}_{Q,(i,z)}\times\mathbf{S}_{Q,(i+\mathbf{d}_\lambda,z)}],\nonumber\\
  \end{eqnarray}
  where $Q$ stands for different spin modes and $\tilde{D}_Q^z(\mathbf{d}_\lambda)$ is the in-plane DMI parameter. The presence of such DMI is consistent with the symmetry of the honeycomb lattice. In momentum space $\delta\mathcal{H}_{\mathbf{k}}=2\eta_{D}\xi_{\mathbf{k}}\mu_z\tau_z$, where $\eta_D=|\tilde{D}^z_Q(\mathbf{d}_\lambda)|/J$.
  Now the system ($\lambda=0$)  has energy
  
  \begin{eqnarray}\label{eq24}
  E_\mathbf{k}^2/(JS)^2&=&|\gamma_\mathbf{k}|^2+4[|\eta_D\xi_\mathbf{k}|\pm\nonumber\\
  &&\sqrt{(2\delta_D\xi_\mathbf{k})^2\cos^2(k_z)+\delta^2\sin^2(k_z)}]^2.
  \end{eqnarray}
  
  Conditions for the existence of Weyl point are $ |\gamma_\mathbf{k}|=0$ and $  \eta_D^2\xi_\mathbf{k}^2=(2\delta_D\xi_\mathbf{k})^2\cos^2(k_z)+\delta^2\sin^2(k_z)$, such that the Weyl nodes lie at $\mathbf{k}_{\parallel}=(0,\pm\frac{4\pi}{3\sqrt{3}})$ and $k_z=\frac{1}{2}\arccos(\frac{3\eta_D^2/2-3\delta_D^2-\delta^2}{3\delta_D^2-\delta^2})$. When $-1<\frac{3\eta_D^2/2-3\delta_D^2-\delta^2}{3\delta_D^2-\delta^2}<1$, there are four-momentum space Weyl nodes originating in the separation of two Dirac cones along $k_z$ direction.
  Similar to Ref.~\cite{PhysRevLett.107.127205}, the system can be manipulated into the Weyl, 3D-mAH, and insulating phases by changing parameters. In parameter space, the insulating and 3D-mAH phases are well separated by the Weyl phase as shown in Fig.~\ref{figure5}, where we identify the 3D-mAH phase by the quantized Chern number ($C=2$ in our model) for arbitrary given $k_z$, i.e., $C=\frac{1}{2\pi}\sum_{E_n<0}\int_{\text{B.Z.}}d\mathbf{k}_\parallel\Omega_{k_x,k_y}^{(n)}(k_z)$ with $\Omega_{k_x,k_y}^{(n)}(k_z)$ being the Berry curvature of bands bellow the gap and $B.Z.$ standing for the 2-D Brillouin zone.
  
  \begin{figure}
  	\centerline{\includegraphics[width=1\columnwidth]{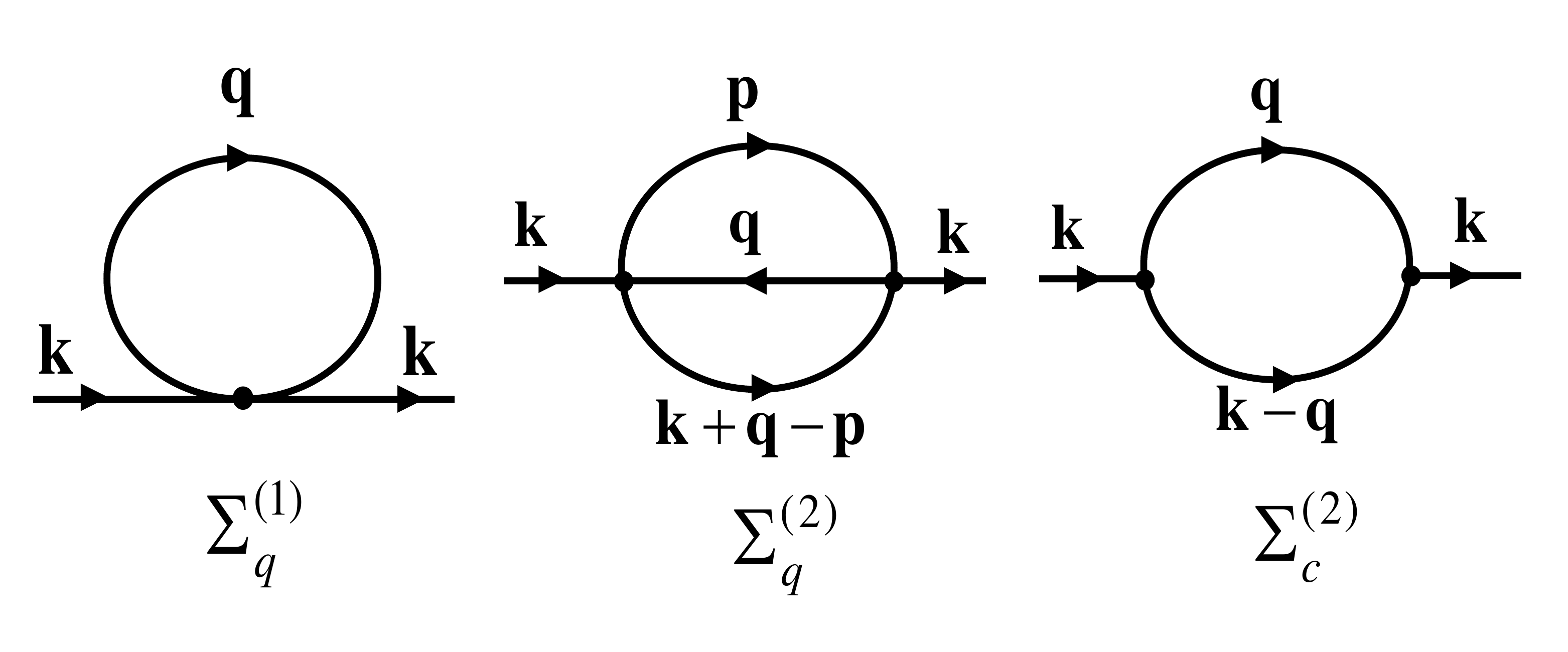}}
  	\protect\caption{(Color online) Left and middle: The self-energy diagrams corresponding to the first- and second-order corrections due to the quartic magnon-magnon interactions. Right: The self-energy diagram corresponding to the cubic magnon-magnon interactions. }\label{figure_Selfenergy}
  \end{figure}
  
  \section{Discussion}
  In this section, we discuss the role of magnon-magnon interaction effects and give possible material candidates for realizations of topological phases of magnons. So far, our discussion has been limited to free magnon systems. It is known that magnon-magnon interactions do not play an important role for a ferromagnetic alignment of spins at low temperatures. In a general case,  magnon-magnon interactions can induce band renormalizations and magnon decay \cite{RevModPhys.85.219}.
It has also been shown that anharmonic terms due to DMI can lead to nonperturbative damping proportional to the strength of DMI in kagome lattice for the spin alignment orthogonal to DMI vectors \cite{PhysRevLett.117.187203}.

We have investigated the role of the above effects in our model by considering the higher-order terms of the Holstein-Primakoff transformation. 
Three diagrams in Fig.~\ref{figure_Selfenergy} contribute to the self-energy where the first two correspond to the quartic term in magnon-magnon interactions and the last one corresponds to the cubic anharmonic interaction. 
According to our analysis, the first two diagrams lead to the self-energy that is proportional to at least the second power of temperature. The effects induced by such diagrams are suppressed at low temperatures since all relevant terms behave in a continuous fashion without singularities. 
As for the third diagram, it is also suppressed by a factor $\propto D^2$ without singularities. The effect of such a diagram completely vanishes for the second model in Appendix~B. For the first model in the main text, we only observe a large contribution when magnetic moments are near orthogonal to DMI vectors. 
This situation can be avoided by tuning the strength of DMI in the model in the main text, in which case the anharmonic contributions do not lead to any singularities. Given nonsingular contributions from all three diagrams, we believe that magnon-magnon interactions cannot hinder topological phases in our models, at least at low temperatures and for typical DMI. 

For realizations of the two models given in the main text and in Appendix B, we suggest to study stacked 2D honeycomb ferromagnets with additional nonmagnetic atoms. From the above discussion it seems that the model in Appendix~B corresponding to $D_{2h}$ point group is better suitable for realizations of the mcTI phase. Among material candidates, one could consider CrI$_3$ van der Waals crystals with honeycomb structure of magnetic atoms \cite{Gong.Li.ea:N2017,Huang.Clark.ea:N2017}. In addition, similar honeycomb magnetic lattices can be realized in transition metal trihalides TX$_3$ (X = F, Cl, Br, and I; T = transition metal) \cite{Jongh:2012}.

 \section{Conclusion}
 In this paper, we constructed a  chiral symmetry-protected topological insulator of magnons in light of the analogous works for electronic and cold-atom systems. In our model, the bulk gap opens due to the presence of DMI. We expect that there could be other magnonic models with mcTI phase and our analysis can facilitate finding other possible realizations. Following the tenfold classification of topological insulators, such models can be characterized by the 3D winding number. We found that the surface  Dirac cone has Rashba-like form, so that the backscattering can be suppressed, which is similar to the surface of the electronic topological insulator. Systems with the broken chiral symmetry at the surface can also be of interest due to a small gap in the surface states and due to appearance of the magnonic Hall response. We showed that the spin Nernst response can be used as a signature of the chiral symmetry breaking at the surface. Finally, we constructed a phase diagram in parameter space, which shows that the system can be tuned between the mcTI, nodal line, 3D-mAH, and Weyl magnon phases. We hope that our work can pave the way for realizations of new topological phases of magnons.

 \section{Acknowledgments}
We thank Vladimir Zyuzin and Kirill Belashchenko for helpful discussions.
This work was supported by the DOE Early Career Award No.DE-SC0014189.

 \bibliographystyle{apsrev}
 \bibliography{MCTI}

\appendix
	
\begin{widetext}

\section{Analysis of possible chiral symmetries for general lattices}

In this Appendix, we explore various possibilities for realizing a chiral symmetry in a system of localized spins. For a  Hamiltonian on a honeycomb lattice with in-plane exchange interactions, we get terms proportional to the following matrices: 

\begin{equation}
\tau_x,\qquad \mu_z\tau_y.
\end{equation}
We further identify possible matrices describing the chiral symmetry,
\begin{equation}
\{\mu_0,\mu_z\}\otimes\tau_z,\qquad \{\mu_x,\mu_y\}\otimes\tau_y.
\end{equation}
We can now write all possible chiral symmetric terms that anticommute with the chiral symmetry. All possibilities are listed in Table~\ref{table1}. For a system of localized spins, we can obtain corresponded hopping terms from exchange interactions and DMI. As a first step, one can construct Dirac magnons and then open a gap with a chiral symmetric perturbation. The minimal model only contains terms that anticommute with each other, but the chiral symmetric perturbations do not necessarily anticommute with the minimal model and can serve to drive the phase transition as discussed in Sec. V.   We note that the presence of the chiral symmetry does not guarantee the mcTI phase and one has to verify the nontrivial topology via winding number calculation. 

The above-mentioned steps can be applied to an arbitrary lattice to obtain other models of mcTIs.

\begin{table}[ht]
	\centering
%	\captionsetup{justification=centering}
	\caption{Symmetry Analysis} 
	\begin{tabular}{c c }
		\hline\hline 
		Chiral Symmetry & Possible Terms \\ [0.5ex] 
		\hline % inserts single horizontal line
		$\tau_z$& $\{\mu_0,\mu_x,\mu_y,\mu_z\}\otimes\{\tau_x,\tau_y\}$\\
		$\mu_z\tau_z$& $\{\mu_x,\mu_y\}\otimes\{\tau_0,\tau_z\}$ $\quad$ $\{\mu_0,\mu_z\}\otimes\{\tau_x,\tau_y\}$\\
		$\mu_x\tau_y$&$\{\mu_y,\mu_z\}\otimes\{\tau_0,\tau_y\}$ $\quad$ $\{\mu_0,\mu_x\}\otimes\{\tau_x,\tau_z\}$\\
		$\mu_y\tau_y$&$\{\mu_x,\mu_z\}\otimes\{\tau_0,\tau_y\}$ $\quad$ $\{\mu_0,\mu_y\}\otimes\{\tau_x,\tau_z\}$\\ [1ex]
		\hline 
	\end{tabular}
	\label{table1}
\end{table}

\section{Model}

In this Appendix, we give more details on mcTI models. First, we show how the DMI is generated in the model we discussed in the main text. Next, we describe a second mcTI model with a different pattern of DMI.
 
\subsection{DMI pattern}
Here, we show how the interlayer DMI can be generated by a nonmagnetic atom in the center of a honeycomb cell. As an example, we calculate the interlayer DMI between $A1$ and $B^\prime1$ spins,

 \begin{figure}[h]
 		\includegraphics[width=0.5\linewidth]{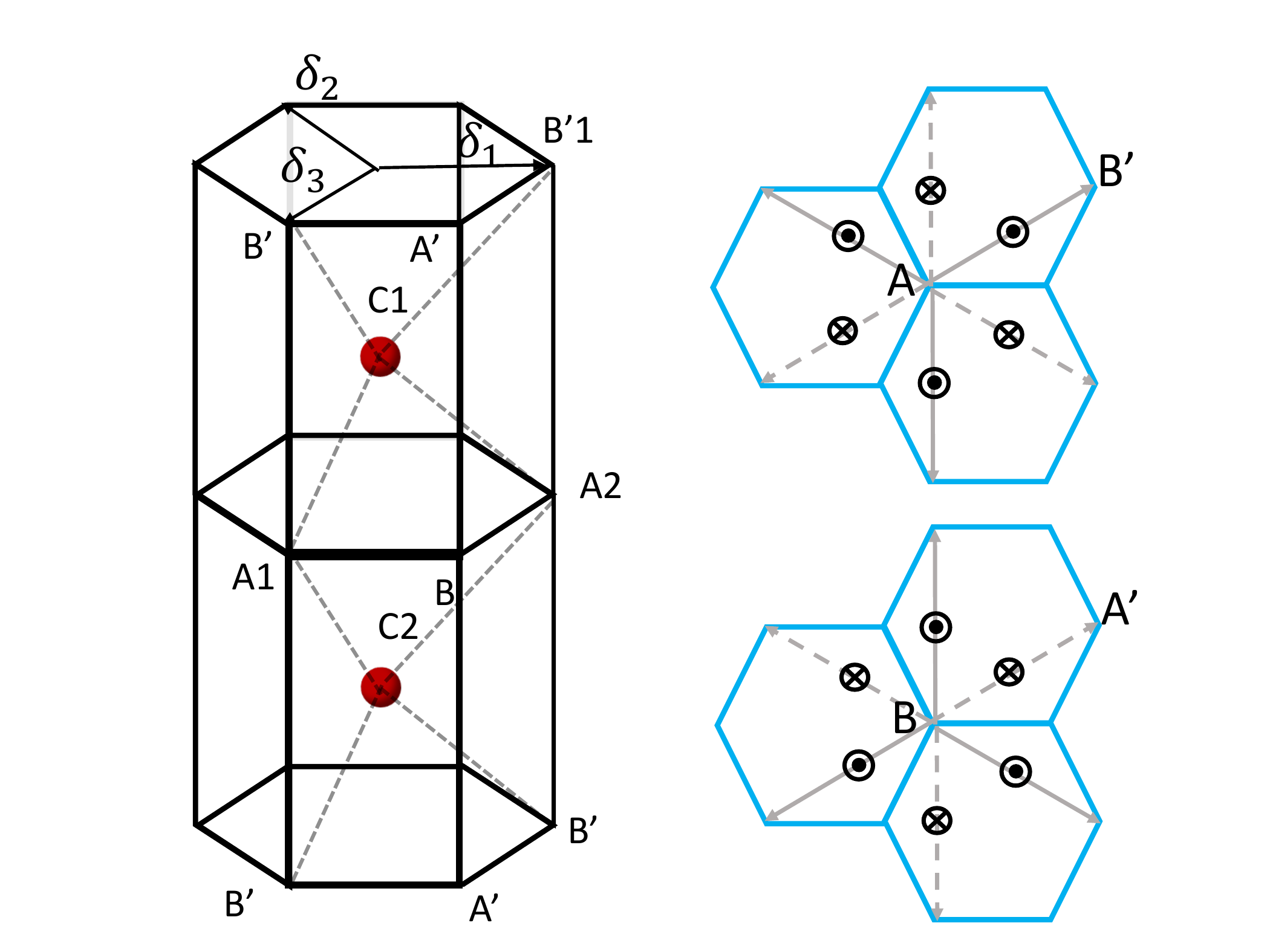}
 	\caption{ The interlayer DMI pattern.}\label{FigA1}
 \end{figure}

\begin{eqnarray}
\overrightarrow{C1A1}=\bm{\delta}_3-\mathbf{c},\qquad \overrightarrow{C1B^\prime1}=\bm{\delta}_1+\mathbf{c},
\end{eqnarray}
where $2\mathbf{c}$ is the vertical interlayer vector, e.g., $\overrightarrow{A2B^\prime1}=2\mathbf{c}$. From the symmetry analysis, the DMI vector between $A1$ and $B^\prime1$ is
\begin{equation}
\mathbf{D}_{A1\rightarrow B^\prime1}=D (\overrightarrow{C1A1}\times\overrightarrow{C1B^\prime1})=D(\mathbf{e}_z+\mathbf{c}\times\bm{\delta}_2).
\end{equation}
The DMI vector $z$-component is $D$, where $D$ is the DMI energy scale. In Fig~\ref{FigA1}, we  give all the DMI
z-component projection, as shown, $B-A^\prime$ and $A-B^\prime$ have opposite sign along the same interaction path vector.\\

\begin{figure}\centerline{\includegraphics[width=0.7\columnwidth]{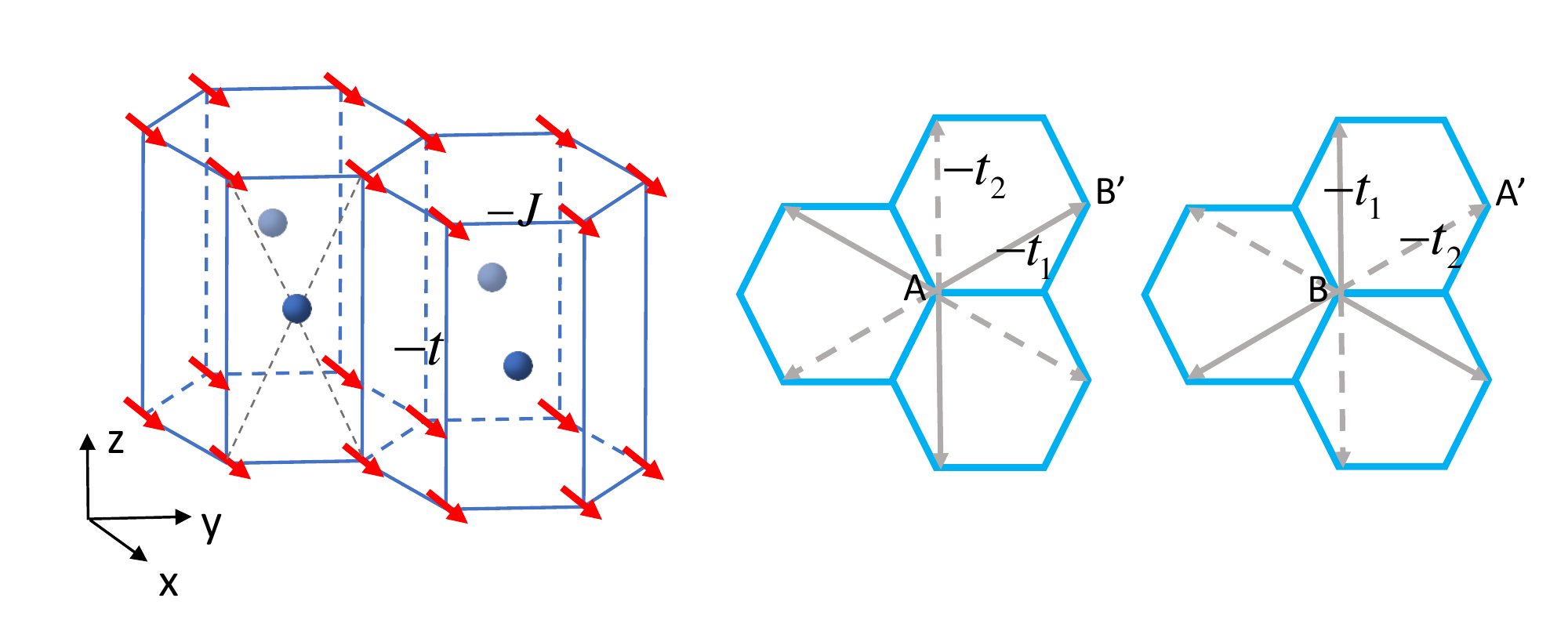}}
	\protect\caption{(Color online) Left: Local spins are pointing in $x$ direction due to applied magnetic field. Nonmagnetic atoms in the face centers generate DMI along the $x$-axis for the vertical bonds. Middle and right: Top view depicts the third-nearest interlayer exchange interactions.}
	\label{figure-Model2}
\end{figure}

\begin{figure}\centerline{\includegraphics[width=0.5\columnwidth]{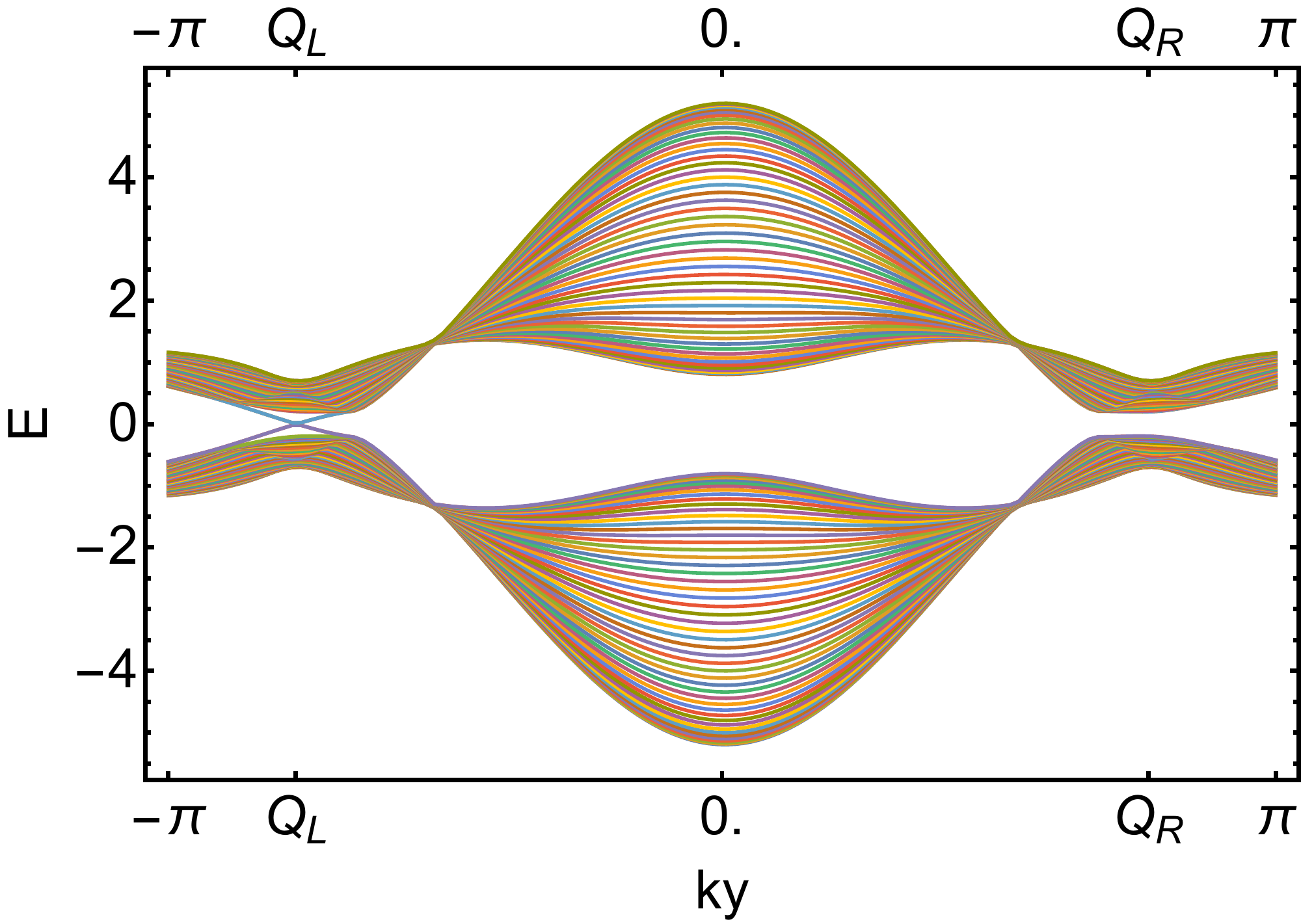}}
	\protect\caption{(Color online) The spectrum of model 2 in a slab geometry shows the presence of surface states. The parameters are $r=0.2$, $\eta=0.1$, $\lambda=0.2$, $\eta_0=0.15$. Here we neglected the boundary effects which shift the position of the surface cone. In principle, this effect can be weakened or even eliminated by an interface with another material.}
	\label{figure-Model2-surface}
\end{figure}

\subsection{Model 2}

Here, we show how a different mcTI model can be realized in a layered honeycomb ferromagnet system. We consider the same lattice structure and labels as in Fig.~\ref{figure1}, but assume that all spins are aligned in $x$ direction, which can be realized by applying an external magnetic field. Instead of putting extra nonmagnetic atoms in the center of unit cell, here we add atoms in the front and back face of each unit cell to generate DMI along vertical interlayer bonds as shown in Fig.~\ref{figure-Model2}. We also need non-uniform third-nearest-neighbor exchange interactions to induce the Dirac cone mass term.  The model Hamiltonian reads
\begin{equation}
H=H_{\text{in}}+H_{\text{inter}}+H_{Z}+H_D+H_{3}^{\text{ex}},
\end{equation}
where
\begin{eqnarray}
H_{\text{in}}&=&-J\sum_{z,i}\sum_{\mu=1}^{3} \mathbf{S}_{A,i}\cdot\mathbf{S}_{B,i+\bm{\delta}_\mu}+\mathbf{S}_{A^\prime,
	i}\cdot\mathbf{S}_{B^\prime,i-\bm{\delta}_\mu},\nonumber\\
H_{\text{inter}}&=&-\sum_{i,z}(t\mathbf{S}_{A,z}\cdot\mathbf{S}_{B^\prime,z+1}+t\mathbf{S}_{A,z}\cdot\mathbf{S}_{B^\prime,z-1}) +(A\rightarrow B,B^\prime\rightarrow A^\prime),\nonumber\\
H_{Z}&=&-\sum_{i,z}\sum_{Q}B_xS_{Q,(i,z)}^x,\nonumber\\
H_{D}&=&\sum_{i,z}\sum_{\delta=\pm 1}\mathbf{D}_{AB^\prime}(\delta)\cdot(\mathbf{S}_{A,z}\times\mathbf{S}_{B,z+\delta})+\mathbf{D}_{BA^\prime}(\delta)\cdot(\mathbf{S}_{B,z}\times\mathbf{S}_{A,z+\delta}),\nonumber\\
H_{3}^{\text{ex}}&=&-\sum_{i,z}\sum_{n=\pm1}\sum_{\mathbf{d}_\lambda}t_1 \mathbf{S}_{A,(i,z)}\cdot\mathbf{S}_{B^\prime,(i+\mathbf{d}_\lambda,z+n)}+t_2\mathbf{S}_{A,(i,z)}\cdot\mathbf{S}_{B^\prime,(i-\mathbf{d}_\lambda,z+n)}+\nonumber\\
&&\{t_1\leftrightarrow t_2, A\rightarrow B, B^\prime\rightarrow A^\prime\}.
\end{eqnarray}
Here, the first two terms coincide with the model in the main text, except that the interlayer nearest exchange interaction has uniform strength. The third term corresponds to the Zeeman interaction with the external magnetic field in $x$ direction. The term $H_D$ represents vertical bond DMI contribution with $\mathbf{D}_{AB^\prime}(\delta)=-\delta D\mathbf{e}_x$ and $\mathbf{D}_{BA^\prime}(\delta)=\delta D\mathbf{e}_x$ ($\delta=\pm 1$). $H_{3}^{\text{ex}}$ stands for the third-nearest-neighbor exchange interaction with staggered exchange strength as shown in Fig.~\ref{figure-Model2}.  After performing the Holstein-Primakoff transformation and the Fourier transformation, the Hamiltonian up to a constant term becomes  
\begin{equation}\label{M3}
\mathcal{H}_\mathbf{k}=-\gamma_{1\mathbf{k}}\tau_x+\gamma_{2\mathbf{k}}\mu_z\tau_y-2r\sin(k_z)\mu_y\tau_y+4\eta\xi_\mathbf{k}\cos(k_z)\mu_x\tau_y+ 2\cos(k_z)(\lambda+2\eta_0\chi_\mathbf{k})\mu_x\tau_x,
\end{equation}
where $r=D/J$, $\lambda=t/J$, $\eta_0=(t_1+t_2)/2J$,  $\eta=(t_1-t_2)/2J$, and $\xi_\mathbf{k}=\sum_{i=1}^3\sin(\mathbf{k}\cdot\mathbf{d}_i)$, $\chi_\mathbf{k}=\sum_{i=1}^3\cos(\mathbf{k}\cdot\mathbf{d}_i)$ with $\mathbf{d}_1=(\frac{3}{2},\frac{\sqrt{3}}{2},0)$,$\mathbf{d}_1=(-\frac{3}{2},\frac{\sqrt{3}}{2},0)$,$\mathbf{d}_1=(0,-\sqrt{3},0)$. First, we consider the extreme case for which $\lambda=\eta_0=0$. The Hamiltonian has the same form as the mcTI model in the main text, i.e., we obtain an effective massive Dirac equation. If we turn on the parameters $\eta_0$ and $\lambda$, they will not immediately break the mcTI phase, similar to the case we discussed in Sec. V in the main text. Specifically, the energy of Eq.~\eqref{M3} is
\begin{eqnarray}
E_\mathbf{k}^2/(JS)^2=[\sqrt{|\gamma_\mathbf{k}|^2+4r^2\sin^2(k_z)}\pm 2|\cos(k_z)(\lambda+2\eta_0\chi_\mathbf{k})|]^2+16\eta^2\xi_\mathbf{k}^2\cos^2(k_z).
\end{eqnarray}
When $\eta=0$, the spectrum is always gapless at two pairs of nodes lying at
$\mathbf{k}_{\text{node}}=(0,\pm\frac{4\pi}{3\sqrt{3}},\pm \arctan(\frac{\lambda+2\eta_0\chi_{\mathbf{Q}_{L/R}}}{|r|}))$. In addition, one needs to assume $\xi_\mathbf{k}=0$ to close the gap, which leads to $k_y=0$ or $\sqrt{3}k_x\pm k_y=0$. For $\mathbf{k}_\parallel^\ast$ satisfying these conditions, the system is gapless at $k_z=\pm\arcsin\sqrt{\frac{4(\lambda+2\eta_0\chi_{\mathbf{k}_\parallel^\ast})^2-|\gamma_{\mathbf{k}_\parallel^\ast}|^2}{4(\lambda+2\eta_0\chi_{\mathbf{k}_\parallel^\ast})^2+4r^2}}$  when $4(\lambda+2\eta_0\chi_{\mathbf{k}_\parallel^\ast})^2-|\gamma_{\mathbf{k}_\parallel^\ast}|^2\geq 0$. When $\max\{4(\lambda+2\eta_0\chi_{\mathbf{k}_\parallel^\ast})^2-|\gamma_{\mathbf{k}_\parallel^\ast}|^2\}<0$, the system is gapped and it is continuously connected with the magnon cTI model with $\eta_0=\lambda=0$ (see Fig.~\ref{figure-Model2-surface}).

\section{Topological invariant}
The Hamiltonian in matrix form is
\begin{eqnarray}
\mathcal{H}_{\mathbf{k}}=\left[
\begin{array}{cccc}
0&-\gamma_{\mathbf{k}}&0&\Delta_\mathbf{k}\\
-\gamma_{-\mathbf{k}}&0&-\Delta_\mathbf{k}&0\\
0&-\Delta_\mathbf{k}^\ast&0&-\gamma_{-\mathbf{k}}\\
\Delta_{\mathbf{k}}^\ast&0&-\gamma_{\mathbf{k}}&0
\end{array}\right],
\end{eqnarray}
where $\Delta_{\mathbf{k}}=-4\delta_D\xi_\mathbf{k}\cos(k_z)+i 2\delta \sin(k_z)$. To transform the matrix to off-diagonal form, we use the transformation operator $U$,
\begin{eqnarray}
U\tau_z\mu_0 U^\dagger= \mu_z\tau_0,
\end{eqnarray}
where 
\begin{equation}
U=\left[
\begin{array}{cccc}
1&0&0&0\\
0&0&1&0\\
0&1&0&0\\
0&0&0&1
\end{array}\right].
\end{equation}
Under the transformation, the Hamiltonian becomes
\begin{eqnarray}
\tilde{\mathcal{H}}_{\mathbf{k}}=U\mathcal{H}_{\mathbf{k}}U^\dagger=\left[
\begin{array}{cc}
0&D_\mathbf{k}\\
D_\mathbf{k}^\dagger&0
\end{array}\right],
\end{eqnarray}
where 
\begin{eqnarray}
D_{\mathbf{k}}=\left[
\begin{array}{cc}
-\gamma_{\mathbf{k}}&\Delta_\mathbf{k}\\
-\Delta_\mathbf{k}^\ast&-\gamma_{-\mathbf{k}}\\
\end{array}\right].
\end{eqnarray}

Assuming that the eigenstates have the form $\psi_a=(\chi_a, \eta_a)^T$, we have
\begin{eqnarray}
\left[
\begin{array}{cc}
0&D_\mathbf{k}\\
D_\mathbf{k}^\dagger&0
\end{array}\right]
\left[\begin{array}{cc}
\chi_a\\
\eta_a
\end{array}\right]=\lambda_a
\left[\begin{array}{cc}
\chi_a\\
\eta_a
\end{array}\right]
\end{eqnarray}
and 
\begin{eqnarray}
\left[
\begin{array}{cc}
D_\mathbf{k}D_\mathbf{k}^\dagger&0\\
0&D_\mathbf{k}^\dagger D_\mathbf{k}
\end{array}\right]
\left[\begin{array}{cc}
\chi_a\\
\eta_a
\end{array}\right]=\lambda_a^2
\left[\begin{array}{cc}
\chi_a\\
\eta_a
\end{array}\right].
\end{eqnarray}
If $D_\mathbf{k}D_\mathbf{k}^\dagger u_a=\lambda_a^2u_a$, with $u_au_a^\dagger=1$ ($a=1,2$), then we obtain  
\begin{eqnarray}
\left[\begin{array}{cc}
\chi_a^\pm\\
\eta_a^\pm
\end{array}\right]=\frac{1}{\sqrt{2}}
\left[\begin{array}{cc}
u_a\\
\pm v_a
\end{array}\right],
\end{eqnarray} 
with 
\begin{equation}
v_a=\frac{1}{\lambda_a}D^\dagger_\mathbf{k}u_a.
\end{equation}
Now, we arrive at the topologically equivalent flat band Hamiltonian
\begin{eqnarray}
Q_\mathbf{k}&=&1-2\sum_{a\in\{ E_a<0\}}|\psi_a\rangle\langle\psi_a|.
\end{eqnarray}
In matrix form, 
\begin{eqnarray}
Q_\mathbf{k}
=\left[\begin{array}{cc}
0&q_\mathbf{k}\\
q^\dagger_\mathbf{k}&0
\end{array}\right],
\end{eqnarray}
with
\begin{eqnarray}
q_\mathbf{k}=\sum_au_av_a^\dagger=\sum_au_au_a^\dagger D_\mathbf{k}\frac{1}{\lambda_a},
\end{eqnarray}
where in our case, $\lambda_a^2=|\gamma_\mathbf{k}|^2+|\Delta_\mathbf{k}|^2$. Consider the negative energy bands (they correspond to the filled bands for electronic systems) and let $\lambda=\sqrt{|\gamma_\mathbf{k}|^2+|\Delta_\mathbf{k}|^2}$, we have 
\begin{equation}
q_\mathbf{k}=\frac{1}{\lambda}D_\mathbf{k}.
\end{equation}
The topology of mcTI is characterized by the 3-D winding number
\begin{eqnarray}
\nu[q]=\int\frac{d^3k}{24\pi^2}\epsilon^{\mu\nu\rho}\text{tr}[(q^{-1}\partial_\mu q)(q^{-1}\partial_\nu q)(q^{-1}\partial_\rho q)].
\end{eqnarray}
We can construct the topologically equivalent Hamiltonian around $\mathbf{Q}_R=(0,\frac{4\pi}{2\sqrt{3}},0)$,
\begin{eqnarray}
\mathcal{H}_{\mathbf{Q}_R+\mathbf{k}}\simeq q_y\tau_x+q_x\mu_z\tau_y+q_z\mu_x\tau_y+m\mu_y\tau_y,
\end{eqnarray}
where $q_x=\frac{3}{2}k_x$, $q_y=\frac{3}{2}k_y$, $q_z=-2\delta k_z$, $m=2\sqrt{3}\delta_D$. It's straight forward to get 
\begin{eqnarray}
D_\mathbf{k}&=&q_y\sigma_0-iq_z\sigma_x-im\sigma_y-iq_x\sigma_z,\nonumber\\
\lambda&=&\sqrt{|\mathbf{q}|^2+m^2}.
\end{eqnarray}
We have 
\begin{eqnarray}
\partial_{q_\mu} q&=&\frac{1}{\lambda}[\partial_{q_\mu}D_\mathbf{k}-\frac{1}{2\lambda^2}(\partial_{q_\mu}\lambda^2)D_\mathbf{k}]=\frac{1}{\lambda^3}(\lambda^2\partial_{q_\mu}D_{\mathbf{k}}-q_\mu D_\mathbf{k}),
\end{eqnarray}
here $\frac{1}{2}\partial_{q_\mu}\lambda^2=q_\mu$, specifically, 
\begin{eqnarray}
\partial_{q_x}q&=&\frac{1}{\lambda^3}[-q_xq_y\sigma_0+iq_xq_z\sigma_x+iq_xm\sigma_y+i(q_x^2-\lambda^2)\sigma_z],\nonumber\\
\partial_{q_y}q&=&\frac{1}{\lambda^3}[(\lambda^2-q_y^2)\sigma_0+iq_yq_z\sigma_x+iq_ym\sigma_y+iq_yq_x\sigma_z],\nonumber\\
\partial_{q_z}q&=&\frac{1}{\lambda^3}[-q_zq_y\sigma_0+i(q_z^2-\lambda^2)\sigma_x+iq_zm\sigma_y+iq_zq_x\sigma_z].
\end{eqnarray}
After some calculation, we obtain
\begin{eqnarray}
\nu_R[q]&=&\int\frac{d^3k}{24\pi^2}\epsilon^{\mu\nu\rho}\text{tr}[(q^\dagger\partial_{k_\mu} q)(q^\dagger\partial_{k_\nu} q)(q^\dagger\partial_{k_\rho} q)]\nonumber\\
&=&-\text{sgn}(\delta)\int\frac{d^3q}{24\pi^2}\epsilon^{\mu\nu\rho}\text{tr}[(q^\dagger\partial_{q_\mu} q)(q^\dagger\partial_{q_\nu} q)(q^\dagger\partial_{q_\rho} q)]\nonumber\\
&=&\text{sgn}(\delta)\int\frac{d^3q}{24\pi^2}\frac{12m}{\lambda^4}\nonumber\\
&=&\text{sgn}(\delta)\frac{m}{2\pi^2}4\pi[-\frac{|\mathbf{q}|}{2(|\mathbf{q}|^2+m^2)}+\frac{1}{2m}\tan^{-1}(\frac{|\mathbf{q}|}{m})]_0^\infty\nonumber\\
&=&\text{sgn}(\delta)\frac{m}{2\pi^2}4\pi\frac{\pi}{4|m|}\nonumber\\
&=&\text{sgn}(\delta)\text{sgn}(m)/2\nonumber\\
&=&\text{sgn}(\delta_D\delta)/2.
\end{eqnarray}
We calculated the case of  $Q_R$ above, for $Q_L$, we only need to replace $q_y\rightarrow-q_y$ and $m\rightarrow-m$, which gives us $\nu_L[q]=\nu_R[q]=\text{sgn}(\delta_D\delta)/2$. After taking all contributions into account, we have the topological invariant
\begin{equation}
\nu[q]=\nu_L[q]+\nu_R[q]=\text{sgn}(\delta_D\delta).
\end{equation}

\section{Surface state}

\subsection{Effective surface Hamiltonian ($\mathbf{k}\cdot\mathbf{p}$ theory)}

We consider the bulk Hamiltonian around $\mathbf{Q}_{R(L)}$ for a system terminated at a honeycomb layer as discussed in the main text. We set $k_x=k_y=0$, keep $k_z$ to second order, and replace it with $-i\partial_z$, 
\begin{eqnarray}
H(z)&=&-A(-i\partial_z)\mu_x\tau_y+M(1-\frac{1}{2}(-i\partial_z)^2)\mu_y\tau_y\nonumber\\
&=&iA\partial_z\mu_x\tau_y+iM(1+\frac{1}{2}\partial_z^2)(\mu_--\mu_+)\tau_y,
\end{eqnarray}
where $A=2\delta$, $M=4\delta_D\xi_{\mathbf{Q}_{R(L)}}$. For the zero-energy surface state,
\begin{equation}\label{D2}
H(z)\psi(z)=0,
\end{equation}
which gives us the form of $\psi(z)$ as 
\begin{eqnarray}\label{D3}
\psi_{1(2)}(z)=\left[\begin{array}{cc}
\bm{\Phi}_{1(2)}\\
\bm{0}
\end{array}\right]e^{\lambda z}.
\end{eqnarray}
Here $\bm{\Phi}_{1(2)}$ is the eigenstate of $\tau_z$ to keep the chiral symmetry, i.e., $(1,0)^T$ and $(0,1)^T$ with eigenvalues $\pm 1$. We plug Eq.~\eqref{D3} into Eq.~\eqref{D2} and obtain
\begin{equation}
A\lambda+M(1+\frac{1}{2}\lambda^2)=0.
\end{equation}
The solution is 
\begin{equation}
\lambda^\pm=-\beta\pm\sqrt{\beta^2-2},
\end{equation}
where $\beta=A/M$, this corresponds to a surface state only if $\beta>0$, i.e., $\text{Re}(\lambda)<0$.  Assuming the boundary condition  $\psi(0)=\psi(\infty)=0$, we obtain two eigenstates:
\begin{eqnarray}
\psi_{1}=N\left(\begin{array}{cccc}
0\\1\\0\\0
\end{array}\right)(e^{\lambda^+z}-e^{\lambda^-z}),\qquad
\psi_{2}=N\left(\begin{array}{cccc}
1\\0\\0\\0
\end{array}\right)(e^{\lambda^+z}-e^{\lambda^-z}).
\end{eqnarray}
Here $N$ is the normalization factor, such that 
\begin{equation}
\int_{0}^{\infty}dzf(z)^\ast f(z)=1,
\end{equation}
with $f(z)=N(e^{\lambda^+z}-e^{\lambda^-z})=2Ne^{-\beta z}\sinh(\sqrt{\beta^2-2}z)$. We can find the normal factor as 
\begin{equation}
N=\frac{\sqrt{\beta^2-2}}{\sqrt{2}|\beta|},
\end{equation}
so that 
\begin{equation}
f(z)=\sqrt{2(1-\frac{2}{\beta^2})}e^{-\beta z}\sinh(\sqrt{\beta^2-2}z)\qquad \text{with} \qquad \beta>0.
\end{equation}
In the vicinity of $(0,\frac{4\pi}{3\sqrt{3}})$, let 
\begin{eqnarray}
H&=&H_{xy}+H(z),\nonumber\\
H_{xy}&=&\frac{3}{2}k_y\tau_x+\frac{3}{2}k_x\mu_z\tau_y.
\end{eqnarray}
It is easy to get 
\begin{eqnarray}
&&\left(
\begin{array}{cc}
\langle\psi_{1}|\\ \langle\psi_{2}|
\end{array}\right)H_{xy}(|\psi_{1}\rangle,|\psi_{2}\rangle)=\frac{3}{2}(k_y\tau_x-k_x\tau_y),\nonumber\\
&&\left(
\begin{array}{cc}
\langle\psi_{1}|\\ \langle\psi_{2}|
\end{array}\right)H(z)(|\psi_{1}\rangle,|\psi_{2}\rangle)=0.
\end{eqnarray}
Therefore, the effective low-energy surface Hamiltonian reads,
\begin{equation}
H_{sur}=\frac{3}{2}(k_y\tau_x-k_x\tau_y)=v_F(\mathbf{k}\times \mathbf{e}_z)\cdot\bm{\tau},
\end{equation}
where $v_F=\frac{3}{2}$. This Hamiltonian possesses spin-momentum locking in the $A$ and $B$ sublattice pseudo-spin space.

\subsection{Surface Hall response}

In order to discuss the Hall response on the surface of mcTI, we start from the gapped surface Hamiltonina (for case $\beta>0$),
\begin{equation}
\mathcal{H}_{\text{sur}}=v_F(\mathbf{k}\times \mathbf{e}_z)\cdot\bm{\tau}+m_s\tau_z.
\end{equation}
We write the Hamiltonian above in a compact form 
\begin{equation}
\mathcal{H}_{sur,\mathbf{k}}=\mathbf{d}\cdot\bm{\tau},
\end{equation}
with $\mathbf{d}=\{v_Fk_y,-v_Fk_x,m_s\}$. The energy and eigenstates are 
\begin{equation}
E_{\mathbf{k},\pm}/JS=\pm d
\end{equation}
and 
\begin{eqnarray}
u_+=\frac{1}{\sqrt{2d(d+d_3)}}\left[
\begin{array}{cc}
d_3+d\\d_1-id_2
\end{array}\right],\qquad
u_-=\frac{1}{\sqrt{2d(d-d_3)}}\left[
\begin{array}{cc}
d_3-d\\d_1-id_2
\end{array}\right],
\end{eqnarray}
where $d=|d|=\sqrt{v_F^2|\mathbf{k}|^2+m_s^2}$. We can define mixed Berry connection as $\bm{\mathcal{A}}_\mathbf{d}^\pm=-i\langle u_\pm|\bm{\nabla}_\mathbf{d}|u_\pm\rangle$. The corresponding Berry curvature is 
\begin{equation}
\bm{\Omega}^\pm(\mathbf{d})=\bm{\nabla}_\mathbf{d}\times  \bm{\mathcal{A}}^\pm(\mathbf{d})=\mp\frac{\mathbf{d}}{2d^3}.
\end{equation}
We use the relation,
\begin{equation}
\Omega^\alpha=\frac{1}{2}\epsilon^{\alpha\mu\nu}\Omega_{\mu\nu},
\end{equation}
with $\Omega_{\mu\nu}=i(\partial_{\mu}\mathcal{A}_{\nu}-\partial_{\nu}\mathcal{A}_{\mu})$. Hence, $\Omega^z=\Omega_{xy}=-\Omega_{yx}$.The spin Nernst response to a temperature gradient is $j^s_y=\alpha_{yx}\nabla_x T$ with
\begin{equation}
\alpha_{yx}=-\frac{k_B}{V}\sum_{\mathbf{k},n}\Omega_{yx}^n(\mathbf{k})c_1(g(\varepsilon_n)),
\end{equation}
where $c_1(x)=(1+x)\ln(1+x)-x\ln x$ and $g(x)=1/(e^{\beta x}-1)^{-1}$. It's easy to check $\Omega^{\pm}_{yx}(\mathbf{k})=v_F^2\Omega^{\pm}_{yx}(\mathbf{d})=-v_F^2\Omega^{\pm,z}(\mathbf{d})=\pm v_F^2\frac{m_s}{2d^3}$. For our system, the response parameter reads,
\begin{eqnarray}
\alpha_{yx}=\frac{k_Bm_s}{2}\int d\mathbf{k}\frac{v_F^2}{d^3}\{c_1[g(\varepsilon_0-d)]-c_1[g(\varepsilon_0+d)]\},
\end{eqnarray} 
where $\varepsilon_0= 3-2\lambda-2\kappa$ and we replaced $\frac{1}{V}\sum_{\mathbf{k}}$ by $\int d\mathbf{k}$. To identify the contribution from the Dirac cone, we introduce a small energy cutoff $\Lambda$ around the Dirac cone, i.e., $\Lambda<\varepsilon_0$. So that we expand $c_1(\varepsilon_0\pm d)$ to the first order of $\beta d$,
\begin{equation}
c_1[g(\varepsilon_0-d)]-c_1[g(\varepsilon_0+d)]=\frac{-\beta^2d\varepsilon_0}{1-\cosh(\beta\varepsilon_0)}+\mathcal{O}[(\beta d)^2].
\end{equation}
Taking the transform $\int d\mathbf{k}=\int_{0}^{\infty}\int_{0}^{2\pi} |\mathbf{k}|d|\mathbf{k}|d\theta=\int_{m_s}^{\infty}\frac{1}{v_F^2}d(d)d\int_{0}^{2\pi}d\theta$, we have
\begin{eqnarray}
\alpha_{yx}&=&\frac{k_B m_s}{2}\int_{0}^{2\pi}d\theta\int_{|m_s|}^\Lambda d(d)\frac{1}{d^2} \{c_1[g(\varepsilon_0-d)]-c_1[g(\varepsilon_0+d)]\}\nonumber\\
&\approx&\frac{\pi k_Bm_s\varepsilon_0\beta^2}{\cosh(\beta\varepsilon_0)-1} \ln(\Lambda/|m_s|).
\end{eqnarray}

\end{widetext}
\end{document}